\documentclass[a4paper,12pt]{elsart}
\usepackage{epsfig}
\usepackage{subfigure}
\newcommand\etc{{\em etc}}
\newcommand\ie{{\bf i.e.}}

\newcommand\ringfinder{{\tt ring-finder}}
\newcommand\MarkovHitRadius{{r_h}}
\newcommand\MarkovRingRadius{{r_R}}
\newcommand\MarkovEpsilon{{\epsilon}}
\newcommand\MarkovAlpha{{\alpha}}

\begin{document}

\begin{frontmatter}



\title{Ring Identification and Pattern Recognition in Ring Imaging
  Cherenkov (RICH) Detectors}


\author{C.G. Lester}

\address{Cavendish Laboratory, University of Cambridge, CB3 0HE, England}

\begin{abstract}
An algorithm for identifying rings in Ring Imaging Cherenkov (RICH)
detectors is described.  The algorithm is necessarily Bayesian and
makes use of a Metropolis-Hastings Markov chain Monte Carlo sampler to
locate the rings.  In particular, the sampler employs a novel proposal
function whose form is responsible for significant speed improvements
over similar methods.  The method is optimised for finding multiple
overlapping rings in detectors which can be modelled well by the LHbC
RICH toy model described herein.
\end{abstract}

\end{frontmatter}

\section{Introduction}

This article describes an algorithm for identifying rings among
photons such as may be observed by Ring Imaging Cherenkov (RICH)
detectors in high energy physics experiments.  The performance of the
algorithm is demonstrated in the context of the LHbC RICH simulation
described in Section~\ref{sec:notes}. (Not the LHCb experiment
\cite{Amato:1998xt,unknown:2000ga})
There are many examples of applications for ring finding pattern
recognition both within high energy particle physics
\cite{Linka:1999gs,Elia:1999sh,Cozza:2001zz,DiBari:2003wy} and without
\cite{ringsInPics,ringsInLogs}.

The first half of the article is entirely devoted to defining what all
ring finders actually {\em are}.  The second half of the article shows
how the idealised ring finder of the first half can be realised by
a {\em real} algorithm, the ``\ringfinder'', to within a good approximation.

We begin with some very simple but very important comments about
pattern recognition in general, and then link these to the specific
case of identifying rings in collections of dots.

The general comments about pattern recognition will make it clear that
{\em meaningful} ring identification can {\em only} take place in the
context of a well defined model for the process believed to have
generated the data containing the ring images.

We therefore go on to describe a model of the way that charged
particles passing through a radiator lead to 
rings of Cherenkov photons being detected on imaging planes.  In
describing this model, we are forced to make explicit our definitions
of rings (both ``reconstructed'' and ``real'') and of hits.  We are
also forced to write down the constituents of the probability
distributions which relate them both.  In terms of these distributions
we are then able to write down {\em the goal} of an ``ideal'' ring
finder in an unambiguous way.  A devout Bayesian could justly say that
there is little or nothing innovative in this first half of the
article -- it being composed largely of definitions and truisms.

The second half of the article
sets out to achieve the goals of the first half, i.e.\ the creation
an actual ring finding algorithm matching the idealised one as closely
as possible.

The method chosen is a Metropolis-Hastings Monte Carlo (MHMC) sampling
of a particular posterior distribution defined in the first half of
the article.  MHMC samplings in ring finders are not new
\cite{Linka:1999gs} but their performance is {\em strongly} dependent
on the choice of the so-called proposal distribution(s) they make use
of (defined later).  Indeed, the {\em only} freedom that one has in
implementing the MHMC sampler, is in the definition of the proposal
distribution(s) to be used -- and so almost all of the second half of
the article is devoted to describing the one used here.

Finally in Section~\ref{sec:secwithreconstructedringsinit} there are
some examples of reconstructed rings.

\subsubsection*{A short note 
concerning common misconceptions about MHMC proposal distributions:}

Among people who have not used MHMC samplers frequently, there is a
common perception that proposal distributions appear to introduce a
level of arbitrariness into the sampler or its results, and there are
often questions or confusion about the manner in which they affect the
results of the sampler.  The short answer is that in the limit of
large times (i.e.\ a large number of samples) the choice of proposal
distributions {\em has no effect at all on the results of the
sampler!}\footnote{In fact parts of the internal mechanism of the MHMC
sampling process described in Section~\ref{sec:methastsamplers} exists
solely for the purpose of {\em removing} any dependence of the results
of the sampler on the choice of proposal distribution.}  However,
differences {\em are} seen after short times (small numbers of
samples).  A clever choice of proposal function allows you to get good
results in a {\em short} time (seconds) whereas a bad choice might
require hours, weeks or even years of CPU time before convergence of
the fit for a single event.  The motivation for choosing good MHMC
proposal functions is thus a desire for {\em efficiency} in the
sampler -- not a desire to introduce some fancy abritrariness or
personal prejudices into it.


\section{Pattern Recognition and Ring Identification}

The single most important thing to recognise when pattern matching is
that:
\begin{quotation}
\bf It is impossible to recognise a pattern of any kind until you have an
idea of what it is you are looking for.
\end{quotation}

\begin{figure}
\centerline{
\subfigure[]{\psfig{file=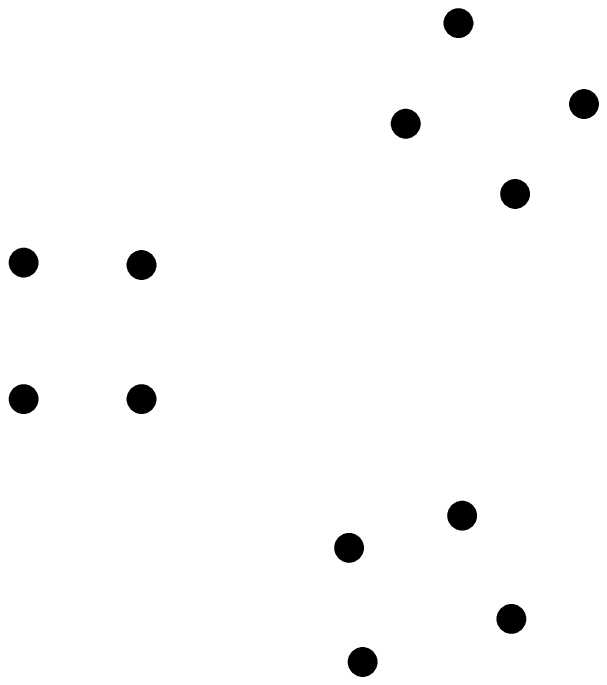,height=3cm}}\qquad\qquad
\subfigure[]{\psfig{file=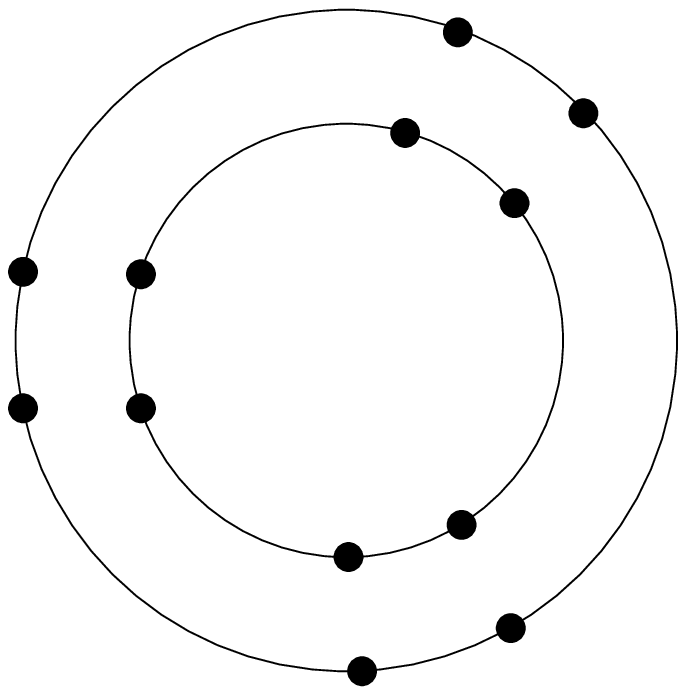,height=3cm}}\qquad\qquad
\subfigure[]{\psfig{file=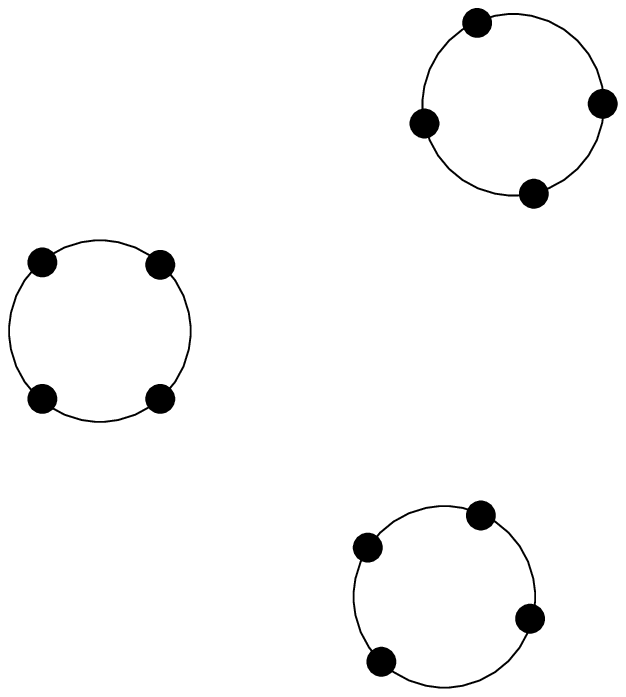,height=3cm}}
}
\caption{What rings should we see in (a)?  Are there two large
  concentric rings as indicated in (b)?  Perhaps there are three small
  rings of equal radii as indicated in (c).}
\label{fig:ambiguousRings}
\end{figure}

To give a simple example from the context of ring finding: What rings
 should a ring finding pattern matcher identify in part (a) of figure
 \ref{fig:ambiguousRings}?

The answer {\em must} depend on what rings we expect to see!

Equivalently, the answer {\em must} depend on the process which is
believed to have lead to the dots being generated in the first place.  If we
were to know {\em without doubt} that the process which generated the
rings which generated the dots in (a) were only capable of generating
large concentric rings, then only (b) is compatible with (a).  If we
were know {\em without doubt} that the process were only capable of
making small rings, then (c) is the only valid interpretation.  If we
know the process could do either, then both (b) and (c) might be
valid, though one might be more likely than the other depending on the
relative probability of each being generated.  Finally, if we were to
know that the process only generated {\em tiny} rings
, then there is yet another way
of interpreting (a), namely that it represents 12 {\em tiny} rings of
radius too small to see.

So any
ring finding pattern matcher must incorporate
knowledge of the {\em process} it assumes lead to the production of
the dots in the first place.

Inevitably it is impossible to know every detail of the process
leading to the generation of the dots, so in practice a ring finding
pattern matcher must at the very minimum have a {\em working model} of
the process that leads to the generation of the dots.

%
%
%
%
%
%
%
%
\section{Rings of Cherenkov Photons in RICH Detectors.}

\label{sec:examplesOfCherenkovRings}
When a charged particle traverses a medium at a speed greater than the
speed of light {\em in that medium}, it emits Cherenkov photons at a
constant angle to its line of flight (but at
uniformly random azimuthal angles).
With an appropriate optical set-up, it may be arranged that
all the 
photons
from a given particle end up striking a screen at points around the
circumference of a ring.  The radius of this ring measures the angle
at which the photons were radiated with respect to the particle's
momentum.  The position of the ring measures the direction in which
the particle was travelling through the medium.  Because the azimuthal
angle of the Cherenkov photons is chosen uniformly, Cherenkov photons
are found uniformly distributed around these rings.\footnote{Subject
to acceptance and optical considerations!}




For the purpose of illustrating the ring-finding technique proposed
herein, we introduce in Section~\ref{sec:notes} a toy model for the
production of hits in an imaginary detector: Lester's Highly basic
Computational (LHbC) RICH simulation.


\section{Modelling for the process of Hit Generation}

\subsection*{Definitions: Rings, Collections of Rings, Hits and Hit Collections}

In the 2-D co-ordinates of a detection plane, a Cherenkov ring $R$ has a
centre ${\bf c}=(x, y)$ and a radius
$\rho$.  Denote a collection of rings by $\bf R$.

When a photon is detected by a RICH detector, or when a
photodetector fires for some other reason, the resulting data-object
will be referred to as a {\em hit}.  For our purposes, the only thing
we need to know about each hit is its position $h$.  The starting
point for the reconstruction of each event is the set ${\bf H} = \{h_i\
|\ i=1,...,n_h\}$ -- the collection of the positions of all of the
hits seen in the event.



\subsection*{Definitions: Low-level and high-level event descriptions}

An event defined as the set of its hit positions $\bf H$ is a
low-level event description. An event defined as a collection of rings $\bf
R$ is a high-level event description.

\subsection*{What the \ringfinder\ is and is not supposed to do}

The purpose of the \ringfinder\ is to make statements about likely
high-level (ring based) descriptions $\bf R$ for an event, given the low-level
(hit based) description $\bf H$ for that event.

The purpose of the \ringfinder\ is {\bf not} to determine the
actual collection of Cherenkov rings that were the cause of the observed
collection of it hits $\bf H$, which will never be known.

Rather it is intended that the \ringfinder\ should {\em sample} from
the space of high-level (ring based) event descriptions
$\bf R$ according to how likely they would appear to have been given
the observed collection of hits $\bf H$.  In other words the \ringfinder\ should supply us with high-level descriptions which ``could
have'' caused the observed data.

\subsection*{Assumptions about the hit-production process}

In constructing a model of the hit production process we assume the
following of the real production process:

We assume that there is an unchanging underlying physical {\bf
M}echanism $M$ which generates events containing an unknown set of
rings $\bf R^{\rm true}$ independently of events produced before or
later.  We assume that there is an unchaning random process $P$,
following on from $M$, according to which a collection of observed
hits $\bf H^{\rm obs}$ is generated from $\bf R^{\rm true}$.  The
random process $P$ is assumed to be known to a reasonable precision
(it is a matter only of known physics and detector response) in
contrast to $M$ which will depend on the type of events the detector
encounters.  Nevertheless, some gross features of $M$ are calculable
(for example detector acceptance may favour central over peripheral
rings) and where they are calculable they may be incorporated.

We assume that $M$ and $P$ can be broken down into parts relating to:


\begin{itemize}
\item
A uniform distribution of Cherenkov photons about the circumference of
the ring,
\item
A poisson distribution for the number of photons likely to be radiated
onto a given ring of radius $\rho$,
\item
Detector resolution,
\item
Backgroud hits coming from random processes unconnected with rings --
for example electronic noise.
\end{itemize}

\label{sec:introducingthedistros}
Formally, all the above information may be encapsulated in two
real-valued functions: the hit-production-model likelihood: 
\begin{equation}
p_P({\bf H^{obs}}|{\bf R^{true}})\label{eq:pOfHGivenR}
\end{equation}
and a probability density function 
\begin{equation}
p_M({\bf R^{true}})\label{eq:pOfR}
\end{equation}
representing the {\em a priori} probability of any particular
configuration of rings, insofar as this can be derived from knowledge
of $M$.  The superscript tags $obs$ and $true$ on $\bf H^{obs}$ and
$\bf R^{true}$ will subsequently be omitted.

The quantity we will ultimately be interested in is $p({\bf R}|{\bf
H})$ which we may obtain from (\ref{eq:pOfHGivenR}) and
(\ref{eq:pOfR}) via Bayes's Theorem:

\begin{equation}
p({\bf R}|{\bf H}) = N({\bf H}) p_P({\bf H}|{\bf R}) p_M({\bf R}),
\label{eq:pOfRGivenH}
\end{equation}

where $N(\bf H)$ is a normalizing constant which we may ignore (set
equal to 1) as we will only be interested in the relative variation of
the left hand side of (\ref{eq:pOfRGivenH}) with respect to $\bf R$
for fixed $\bf H$.  No further mention will be made of $N(\bf H)$.

Note that the dimension of $R$ is three times the number of rings it
contains as each ring is defined by a $2$-dimensional centre and a
$1$-dimensional radius.  A typical event contains order 10 rings,
and so $p({\bf R}|{\bf H})$ is typically a function of order 30
dimensions.  It will not therefore be possible to plot $p({\bf R}|{\bf
H})$.  However, we can do the next best thing: we can use Monte Carlo
methods to sample from it.

The set of high-level descriptions $R$ which we will draw from $p({\bf
R}|{\bf H})$ will represent the most reasonable guesses we can make for
$\bf R^{\rm true}$ given $\bf H$.

Note that we do not make any attempt to ``maximise'' $p({\bf R}|{\bf
H})$.  We are not interested in where this function is a
maximum,\footnote{Note: It may seem strange that we are not interested
in the maximum of $p({\bf R}|{\bf H})$ given the large amount of the
literature devoted to maximum-likelihood analyses.  But bear in mind
that (1) the position of the peak is not invariant under
reparametrisations of the space, and (2) this is a high-dimensional
problem.  In high-dimensional problems the vicinity of the maximum 
is often either only a tiny part of the ``typical set'' (the region
containing most of the {\em probability mass} or else is not even
part of the typical set at all!  See \cite{MacKay} for detailed
discussion.} though in some sense the sampling is likely to be
localised near the maximum.


Note that the algorithm described herein is ``trackless'' acting
only on the hits generated by LHbC RICH simulation of Section~\ref{sec:notes}.
Trackless ring-finding algorithms have been proposed in the past,
however. Reference \cite{Linka:1999gs} only came to the attention of
the authors two years after the algorithm defined herein was
implemented.  The algorithm of \cite{Linka:1999gs}, though
independently concieved, has much in common with the one described
here, and has much to commend it.  Both methods use a Bayesian
approach, implement a similar detector model\footnote{Though
\cite{Linka:1999gs} describes a detector with analouge hit information
rather than digital as in this paper.} and explore the space of
possible ring-configurations with a Markov chain Monte Carlo.  The
details of each of the algorithms' Metropolis-Hastings proposal
functions differ very significantly, however, and it is this
difference which the authors believes accounts for the significant
improvements in efficiency (speed) and performance of the fitter
described herein in situations of high ring multiplicity.

\subsection*{Summary of the ring-finding method}

The description of the ring-finding method in the preceding sections
may be summarised in two steps as follows:
\begin{itemize}
\item
\label{step:defdists}
Define the types of rings you want to find by calculating the
specific forms of the distributions in Equations~(\ref{eq:pOfHGivenR})
and (\ref{eq:pOfR}) which are relevant to the production model (in
this case LHbC RICH simulation of Section~\ref{sec:notes}.)
\item
Sample from the resulting posterior distribution $p({\bf R}|{\bf H})$
(Equation~(\ref{eq:pOfRGivenH})).
\end{itemize}

Of the two steps above, the first one is by far the simplest.  It
leaves almost no scope for flexibility or creativity.  Either the
calculated distributions are or are not a fair representation of the
mechanism of ring production, detector response and hit generation for
the problem in question.  Newer models of the process or the detector
can be switched in at short notice for comparison with older models,
with no significant impact on other parts of the \ringfinder.  The
particular forms where were used here are described in
Section~\ref{sec:notes}.

It is the second of the steps above that is the hardest and is the
part which will take up most of the rest of the discussion.

It will become clear later that while it would be possible to
implement a ``general'' second step,\footnote{{\em i.e.} a second step
that is not specifically tailored to the problem in question} it is
almost certain that such a method would be hopelessly inefficient and
completely unusable.  We will therefore always discuss the second step
in the context of the particular sort of ring finding that is required
by the LHbC RICH simulation of Section~\ref{sec:notes}.

\subsubsection*{Why is a general method of plotting $p({\bf R}|{\bf H})$ hopelessly
inefficient?}

It has already been mentioned that $p({\bf R}|{\bf H})$ as a function
of $R$ is a function of around $30$-dimensions, and we know that we
are interested in discovering where in this space the bulk of the
probability lies.  The large dimensionality of the space precludes
simply plotting the density itself, and suggests that a
sampling or explorative method is instead required.

If the space had only one probability maximum\footnote{\ie\ if the
space had no local maxima other than the global maximum} then
established techniques (such as steepest descent \etc) could be
employed to find out where the ``centre'' of the distribution was, and
then the simplex or other multi-point methods could probably be used
to explore the bulk region.  Unfortunately, the space is actually
packed full of local maxima separated by regions of improbability, and
so these methods would very quickly get irretrievably stuck in poor
local maxima.

Take the example shown in Figure~\ref{fig:ambiguousRings}.  There is
no way to slowly transform the two fitted rings of (b) into any two of
the three rings in (c) which does not involve passing through a huge
region inbetween in which the rings would represent a {\em terrible}
fit to the observed hits.

The only real solution appears to be to sample the space using a
custom Markov Chain Monte Carlo (MCMC) sampling method -- one that has
built into it an understanding of the type of space it is trying to
explore, and an ability to make sensible guesses as to the locations
of distant isolated local maxima.

This is all necessary to improve the {\em efficiency} of the \ringfinder\ to the point at which it can become {\em useful}.  Wherever
possible, the choice and design of the custom Markov sampler should
not affect the answers that are reached, only the time it takes to
reach them.\footnote{By way of an example: the simplest possible
Markov Chain sampler would probably be one using the Metropolis Method
with a flat proposal function.  This method would indeed sample the
space exactly as required, but in 30 dimensions you would probably
have to wait an unfeasable ${100}^{30}$ iterations before the
distribution converged on the right answer (assuming 1\% scan
granularity).
}

The approach adopted herein was to use a Metropolis-Hastings sampler
with a proposal distribution tailored specifically to the sorts of
hits seen in the LHbC RICH simulation of Section~\ref{sec:notes}.

\section{Metropolis-Hastings Samplers and Proposal Distributions}

\label{sec:methastsamplers}
 
A full review of Sampling Theory and Markov Chain Sampling techniques
is beyond the scope of this article.   What follows only describes
the bare minimum needed to implement the sampler used by the
\ringfinder.  No attempt is made to explain why the described
procedure does indeed perform a statistically correct sampling --
though references to items relevant in the literature are given.

In general, a Metropolis-Hastings sampler
\cite{MacKay,Metropolis,Hastings} samples a sequence of points
$\{x_i\}$ from some space $X$ on which a target probability
distribution $p(x)$ has been defined.  Suppose $n$ points ${x_1, ... ,
x_n}$ have already been sampled.  The next point $x_{n+1}$ is sampled
as follows.  A {\em proposed} location $w_{n+1}$ for the next point is
drawn from a ``proposal distribution'' $Q(w|x)$ with, in this case,
$x=x_n$.
The only two requirements of the proposal distribution are (1) that it
be easy to draw uncorrelated samples from $Q$, and (2) that it be
possible to calculate $Q(w|x)$ up to an abritrary constant factor.  A
dimensionless random number $\rho$ is then drawn uniformly from the
interval $[0,1]$.  If $\rho$ is found to be less than $\rho^{\rm max}
= \frac{p(w_{n+1})Q(x_n|w_{n+1})}{p(x_n)Q(w_{n+1}|x_n)}$, then the
proposal $w_{n+1}$ is accepted, and $x_{n+1}$ is set equal to
$w_{n+1}$.  Otherwise, $x_{n+1}$ is set equal to $x_n$.

In this particular paper, the $p(x)$ mentioned above will be the
$p({\bf R}|{\bf H})$ first seen in Equation~(\ref{eq:pOfRGivenH}), and
so $X$ will be the space of high-level event descritons $\{\bf R\}$
(the space of ring hypotheses).

Note that a Metropolis-Hastings sampler does not in general produce
uncorrelated samples (in fact there is a very high chance that two
more more neighbouring samples may actually be identical!) however as
the number of samples tends to infinity, the resultant set of samples
may be treated as if they were the result of an uncorrelated sampling
process.

The art of creating the Metropolis-Hastings sampler that is most
suited to a given target distribution is equivalent to finding the
best proposal distribution $Q(w|x)$ for the problem.\footnote{Note
that if $Q(w|x)$ were chosen to be $p(w)$ then $\rho^{\rm max}$ would
always be exactly 1, every proposed point $w$ would thus be accepted,
and all samples would be completely independent samples from $Q$ and
thus would also be completely independent samples from the target
distribution $p$.  This is in some sense the ``optimal'' $Q$.  But one
of the requirements of $Q$ was that we should be able to sample from
it, and if we were able to sample from $p(w)$ directly we would have
no need of the Metropolis-Hastings method to sample from $p(x)$.  So
in practice the optimal $Q$ is (a) one from which it {\em is} possible
to draw samples, but (b) is nonetheless ``as close as possible'' to
the target distribution $p$.}  There is considerable scope for
creativity in the construction of good proposal distributions for
particular problems.  Despite much development, the proposal
distribution described herein (outlined in
Section~\ref{sec:ourproposaldist} with details in
Section~\ref{sec:notes}) is unlikely to be optimal for the LHbC RICH
simulation of Section~\ref{sec:notes}.
Further development of the \ringfinder's proposal distribution is the
single most important objective in any attempt to improve ring finding
performance -- judged according to how long you must wait before the
samples are representative of the whole distribution.

\section{An MHMC proposal distribution suitable for ring-finding in
  LHbC RICH simulation of Section~\ref{sec:notes}}

\label{sec:ourproposaldist}

The proposal distribution used in the \ringfinder\ is best
described algorithmically.  At the top level, it can:
\begin{enumerate}
\item
propose the {\bf addition} of a new ring-hypothesis to the current set of ring-hypotheses,
\item
propose the {\bf removal} of a ring-hypothesis from the current set of ring-hypotheses, or
\item
propose a small {\bf alteration} to one or more of the existing ring-hypotheses. 
\end{enumerate}
By tuning the relative probabilities with which the above options are
chosen, one can try to maximise the efficiency of the
sampler.\footnote{In the context of this document, the efficiency of
the sampler is always defined as the reciprocal of the inefficiency of
the sampler, which itself is defined as the proportion of proposals
which are wasted.  Wasted proposals are ones which are rejected by the
Metropolis algorithm, thus causing the current point to be re-visited
as the next point of the sampling.}  A crude attempt at optimisation
was done by hand but there will be scope for improvement.  At the time
of writing, the values in use were as follows.  Propose an alteration with
probability 0.2.  If not making such a proposal, propose a circle
addition with probability 0.6.  Otherwise propose a circle removal.


\subsection*{Alterations to ring hypotheses}

The ``alteration'' option allows the positions and radii of previously
proposed rings to be ``fine tuned'' to better reflect their likely
locations in the light of neighbouring rings etc.  

\begin{figure}
\centerline{
\subfigure[]{\psfig{file=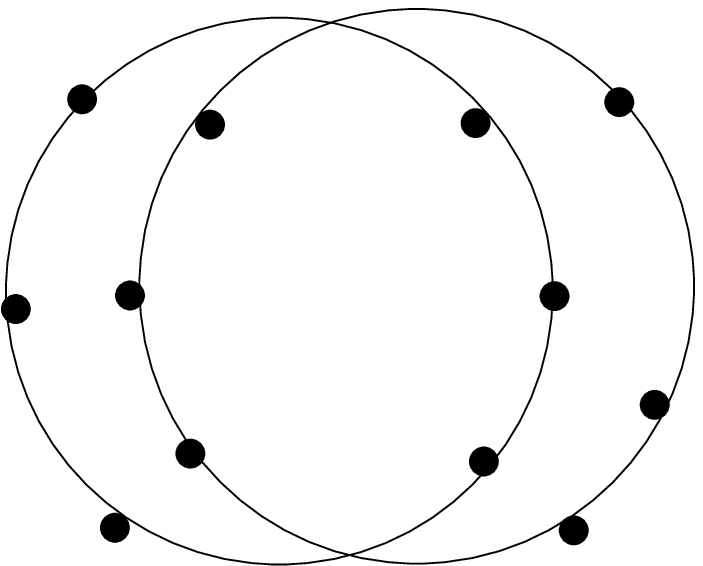,height=3cm}}\qquad\qquad
\subfigure[]{\psfig{file=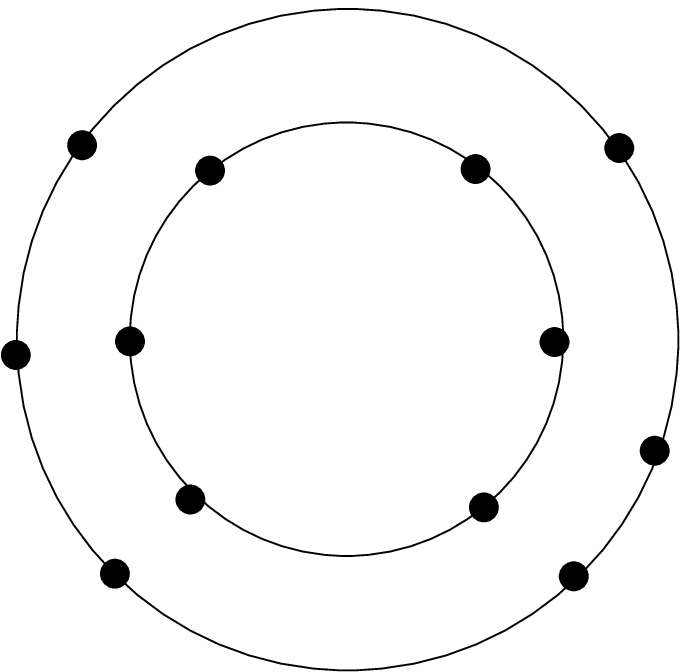,height=3cm}}
}
\caption{Part (a) shows a common example of a mis-fitting.  In
  this example, the fit should really have been as shown in part (b).}
\label{fig:ambiguousRings2}
\end{figure}

Most of the time, fine tuning is most efficient if it involves
perturbing the position and size of only one ring.  More often than
not, a perturbation to any one ring results in a ring which fits worse
than the original, and so as the number of simultaneously
perturbed rings grows, the chance that the proposal will be
accepted by the Metropolis algorithm diminishes exponentially.

Nevertheless, there is a common situtation depicted in
Figure~\ref{fig:ambiguousRings2}, in which it is beneficial to try to
perturb two rings at once in order to pull a bad fit out of a false
minimum.  In cases like this, it is unrealistic to expect the
ring-finder to switch from \ref{fig:ambiguousRings2}(a) to
\ref{fig:ambiguousRings2}(b) by successive perturbations of any one
ring, or by removal and subsequent reinstatement of both rings, as the
potential barrier to this (the poor quality of the intermediate fits)
would lead to exceptionally long equilibrium times.

To take this and similar situations into account, the number of ring
hypotheses which are the subject of modification in a given
``alteration'' is itself chosen at random from a distribution which
favours single rings over pairs and pairs over triplets \etc.  The
precise choice of this distribution is again something which may be
tweaked to increase the efficiency of the sampler.  At the time of
writing, the number $n$ of ring hypothesis to be altered
in a given ``alteration'' (out of a total number of ring hypotheses
$N$) was selected with a probability proportional to $1/n$.  It is
likely that a better choice leading to a more efficient sampler could
be found.

\begin{figure}
\centerline{\psfig{file=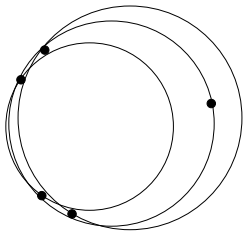,height=3cm}}
\caption{All three of the above rings fit the four points on the left
  hand side reasonably well.  If the ``current'' ring hypothesis were
  to be the largest or the smallest ring above, one would like to be
  able to shrink or grow the ring about the well fitted points until
  the remaining hit on the right hand side were to also be fitted.  No
  such ``correlated smearining'' is imlemented in the current code, so
  the same effect has to be reached by a $3$-dimensional random walk,
  at the cost of efficiency .}
\label{fig:correlatedSmearings}
\end{figure}

Once the number $n$ and identity of ring hypotheses to be perturbed
has been chosen, the perturbation of each ring hypothesis is performed
by independent symmetrical Gaussian smearings of each of the three
ring coordinates (centre-$x$, centre-$y$ and radius).  In each case
the width of the smearing is equal to 10\% of the average radius of a
typical ring.  There is again scope for
optimising this mechanism in order to make the sampler more efficient.
In particular, when more than one ring is simultaneously modified, it
would make sense to allow smearings to correlate between the rings.
This might help to more efficiently remove mis-fits like that shown in
Figure~\ref{fig:ambiguousRings2}.  Also it might be an idea to
consider correlated smearings within the three parameters of a single
ring so that (for example) one could leave the best-fitted parts of
the ring as unaltered as possible (see
Figure~\ref{fig:correlatedSmearings}) while allowing the less well
constained parts of the ring to move as much as possible.

\subsection*{Addition of new and removal of old ring hypotheses}

Given that the decision to insert-or-remove a ring has been
made, the deletion is proposed with probabiity $0.4$ and insertion
with probability $0.6$.

\subsubsection*{Removal of old ring hypotheses}

Once scheduled, removal of a ring hypothesis is as simple as it
sounds.  The only thing worth mentioning is that the removed
hypothesis is not thrown into a black hole and lost forever.  Instead
it gets pushed onto a stack of ``ring hypotheses which have been useful in
the past''.  The use of this stack is discussed later.

\subsubsection*{Addition of new ring hypotheses}

If the three coordinates (centre-$x$, centre-$y$ and radius) of a ring
hypothesis are drawn at random from their whole-experiment average
distributions, it is highly unlikely that the resulting ring will
correspond to a ring in the data.  There may only be $20$ to $50$ real
rings in an event, but there are of the order of $100*100*20$
distinguishable ring hypothesis you could make.  A ring hypothesis
drawn at random thus has roughly only a one in ten thousand chance of
being close-to-useful.  To improve the efficiency of the sampler,
proposals for new rings must have a better means of making
suggestions.

The approach taken in the \ringfinder\ is to try to seed ring
suggestions from groups of three hits.  Again, it is not good enough
to choose just any three hits at random, as a typical event can have
upwards of 300 hits (say 15 hits per ring) so the chance of three hits
drawn at random coming from the same ring is of the order of
$(15/300)^2 = 1/400$ which is still too small to be useful.  Instead
the three points are chosen in a correlated manner termed the ``three
hit selection method'':

\subsubsection*{The three hit selection method}

\begin{figure} 
\centerline{
\psfig{file=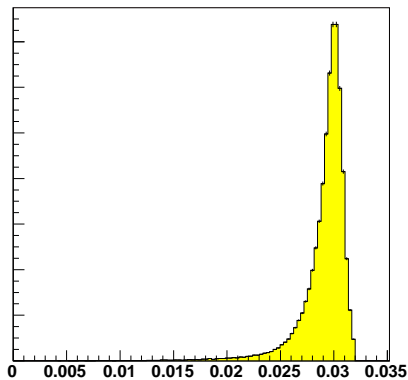,height=6cm}}
\caption{The ring radius distribution (horizontal axis in radians)
  used in the LHbC RICH simulation of Section~\ref{sec:notes}.
  The functional form is shown in Equation~\ref{eq:radiusdistroeq}.
\label{fig:ringRadii}}
\end{figure}

First one of the hits in the event is chosen at random.  Then all
other hits in turn are compared with the first hit.  For each hit, the
likelihood that it (given no other information) is in the same circle
as the first hit is calculated.  This may be done purely on the basis
of the knowledge of the whole-experiment ring radius distribution
(Figure~\ref{fig:ringRadii}) and a little numerical
integration.\footnote{It might be objected that the whole-experiment
ring radius distribution is not known (except from Monte Carlo event
generation) before the experiment turns on, and can only be measured in
a RICH detector, and so training a RICH new-ring proposal distribution on
the basis of Monte Carlo predictions will introduce some sort of bias
into the \ringfinder.  Fortunately this is not a worry, as once again
the purpose of the modified proposal distributions is not to change
the answer,
only to reach the answer more efficiently.  If the Monte Carlo data
were not to match the experimental data very well, that would only
make this proposal distribution a little bit more inefficient than
intended ... it would not invalidate the result otherwise.}  Once all
such likelihoods have been calculated, one of these hits is chosen
(with a probability proportional to its likelihood) to join the first.
By this stage we have selected two hits which have a reasonable
probability of being in the same ring.  We now need to choose a third.
A similar procedure is followed as before.  All other hits are
compared with the first two, and the likelihood that (given no other
information) they are in the same ring as the first two is calculated,
and one of the hits is then chosen to join the first two with a
probility proportional to its likelihood of being in the same ring.
Again this depends only on knowledge of the whole-experiment ring
radius distribution.  Having selected three points likely to be in the
same ring, the ring passing through all three hits becomes the
proposal which is offered to the Metropolis method for approval or
rejection.

\begin{figure} 
\centerline{
\psfig{file=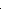,height=6cm}
\psfig{file=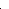,height=6cm}} 
\caption{Example of 100 new rings {\em proposed} by the ``three hit
  selection method'' for consideration by the MHMC for possibile
  inclusion in the final fit.  The hits used to seed the proposal
  rings are
  visible as small back circles both superimposed on the proposals
  (left) and on their own (right). 
\label{fig:examplesOfThreeHitProposals}}
\end{figure}

Figure~\ref{fig:examplesOfThreeHitProposals} shows 100 circles {\em
proposed} by the three hit selection method for an example event.  As
desired they are concentrated mostly in areas where rings appear to be
-- not much time is being wasted proposing wildly unrealistic circles.
Note that not all of the proposed circles will be accepted by the MHMC
algorithm.  Proposal and acceptance are quite different things within
the MHMC sampler.

Of course, badly distorted rings and rings with fewer than three hits
on them will never be seeded this way, so the above prescription is
applied only 90\% of the time.  The remaining 10\% of the time the
proposal function falls back on a naughty trick -- it suggests a
``reverse ring addition''.  A ``reverse ring addition'' is the
popping off and subsequent re-use of the top-most ring in the stack of
``ring hypotheses which have been useful in the past''\footnote{See
``Removal of old ring hypotheses'' at the start of this section} as
the new-ring proposal.  In the event that the stack is empty, the
proposal method falls back to the most basic of the new ring proposal
mechanisms described at the beginning of this section, even though it
is very unlikely to lead to much success.  A ``reverse ring addition''
is not strictly a valid action in the context of the Metropolis method
as it breaks the principle of detailed balance.  However in practical
terms it has proved to be a beneficial thing to have inside the
sampler, and it does not seem to break the principle of detailed
balance enough to cause any obvious problems.  The ``reverse ring
addition'' mechanism allows the \ringfinder\ to be a little more
aggressive about throwing ring hypotheses away than it would otherwise
be able to be -- if it gets second thoughts about a disposal, it
effectively has a chance to change its mind and recover a ring
hypothesis that it had thrown away earlier.

\section{Results}

\begin{figure}
\subfigure[event 1]{
\psfig{file=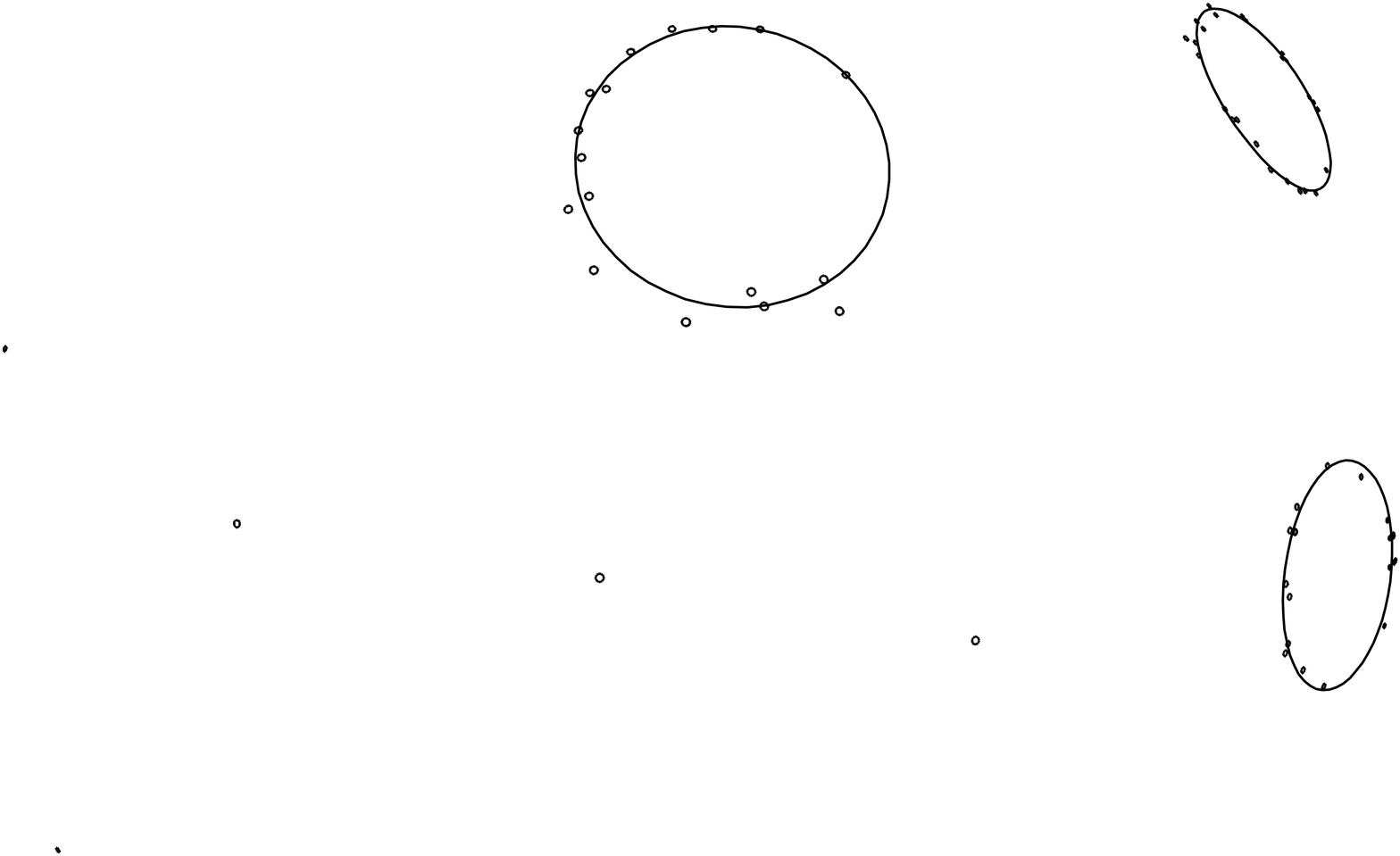,height=3cm}
\psfig{file=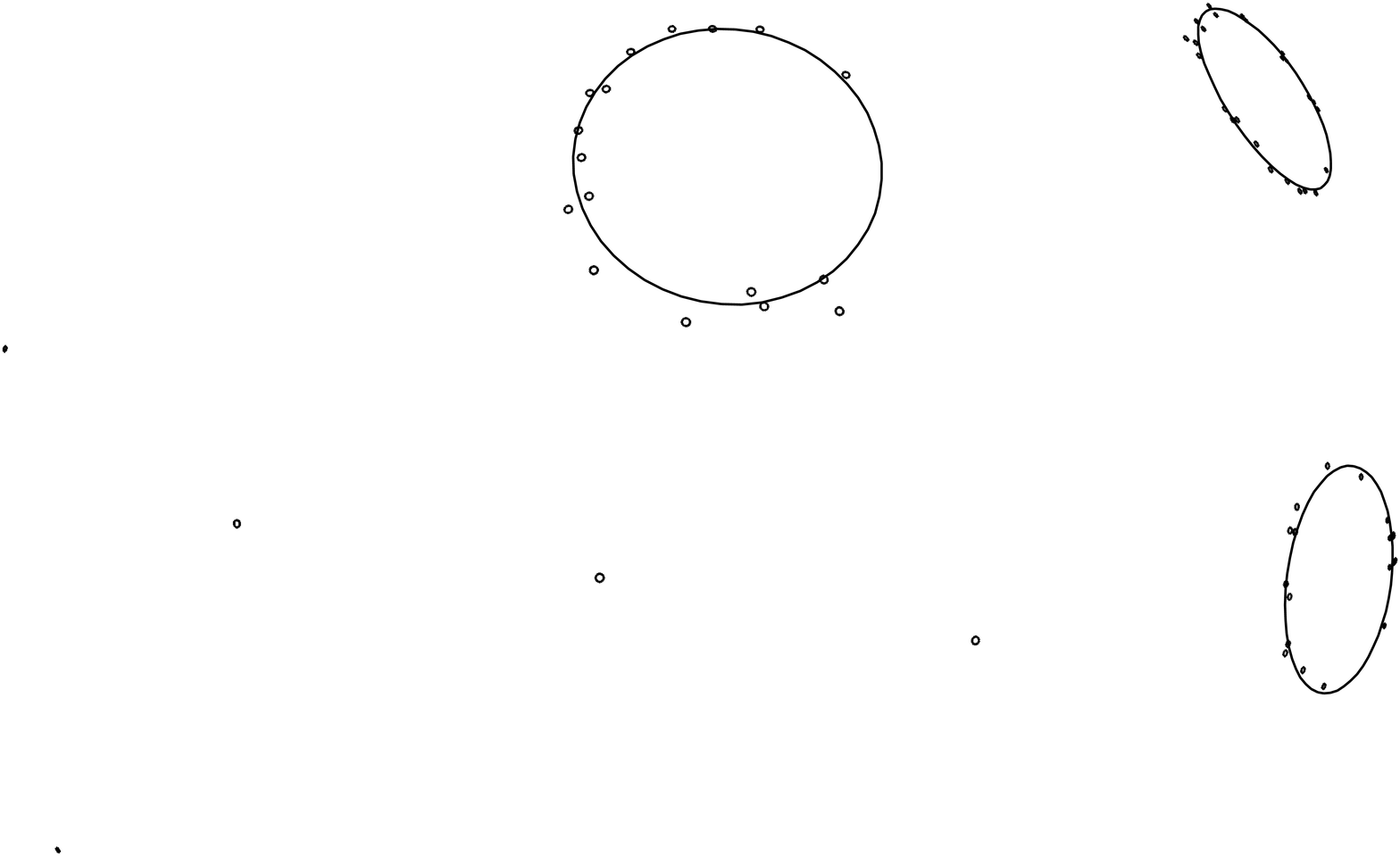,height=3cm}
}
\subfigure[event 2]{
\psfig{file=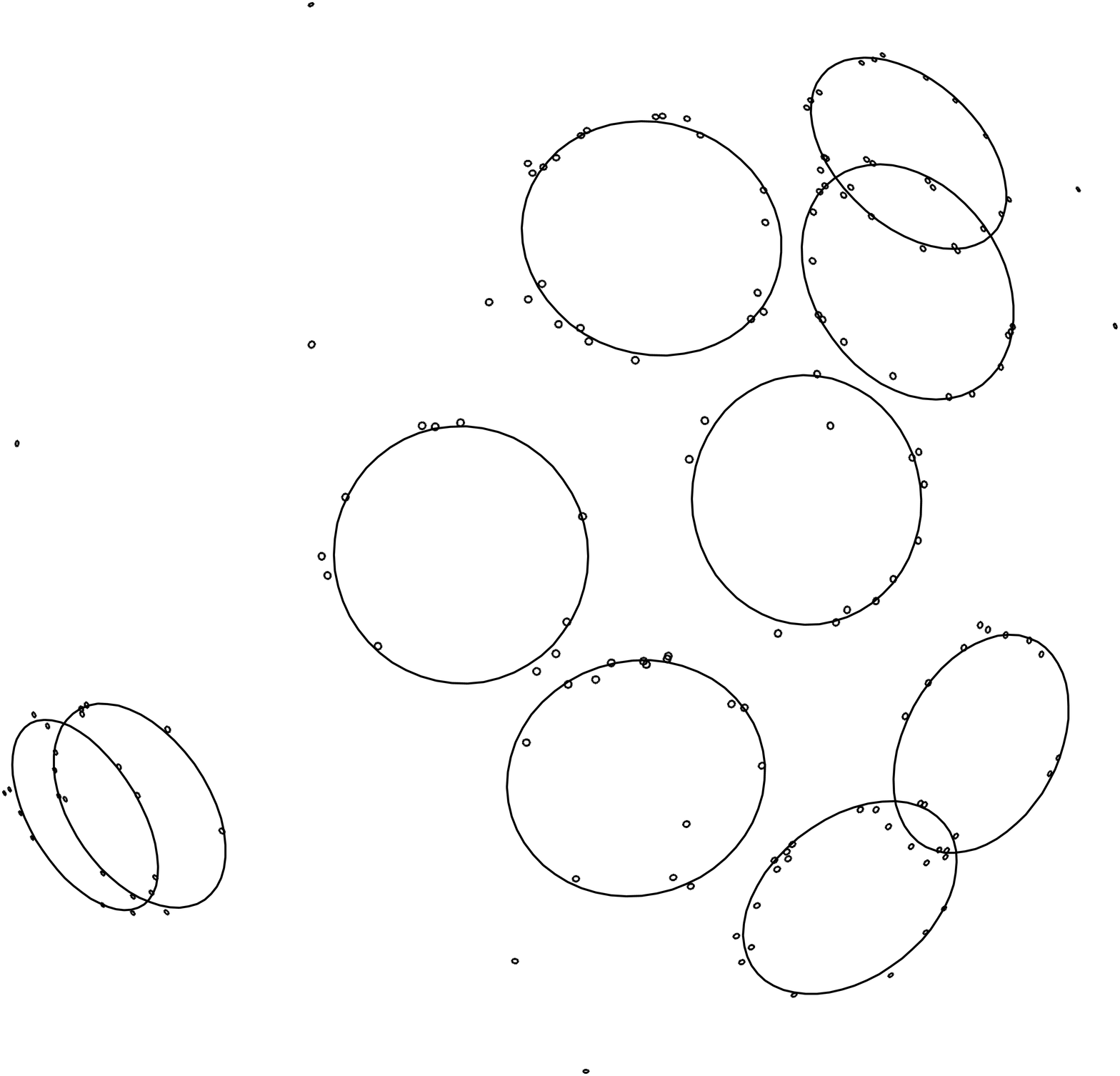,height=3cm}
\psfig{file=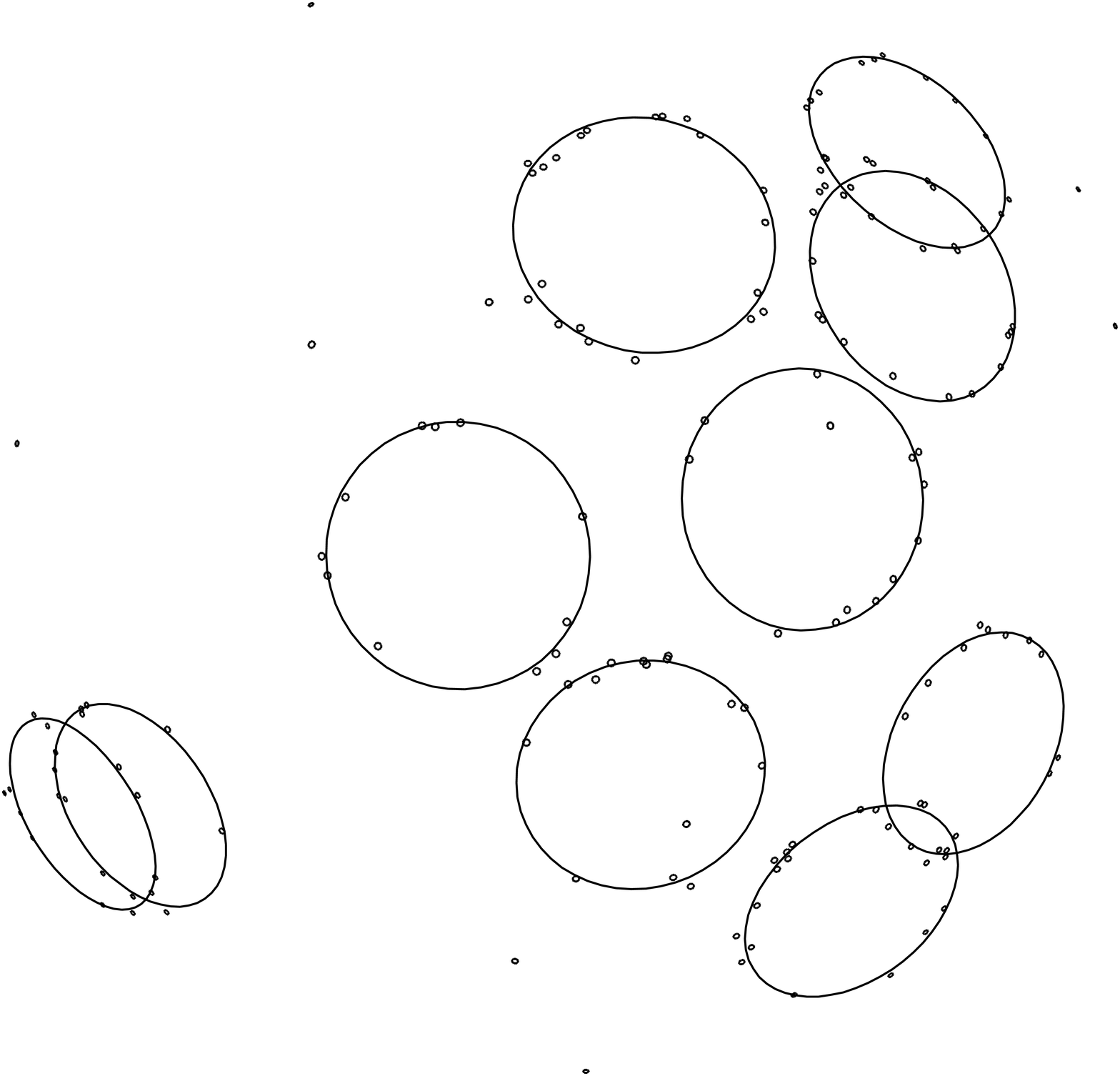,height=3cm}
}
\subfigure[event 3]{
\psfig{file=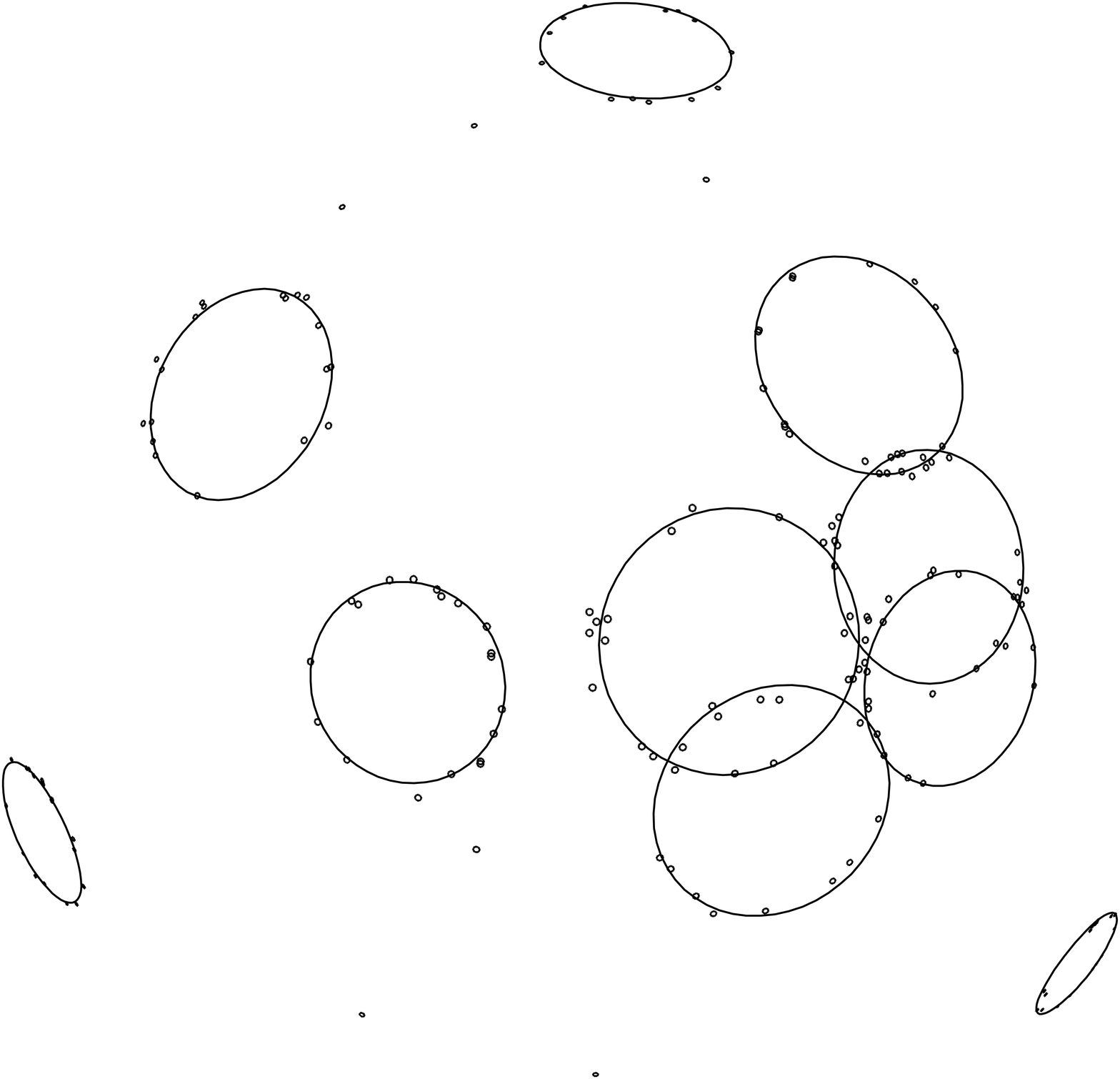,height=3cm}
\psfig{file=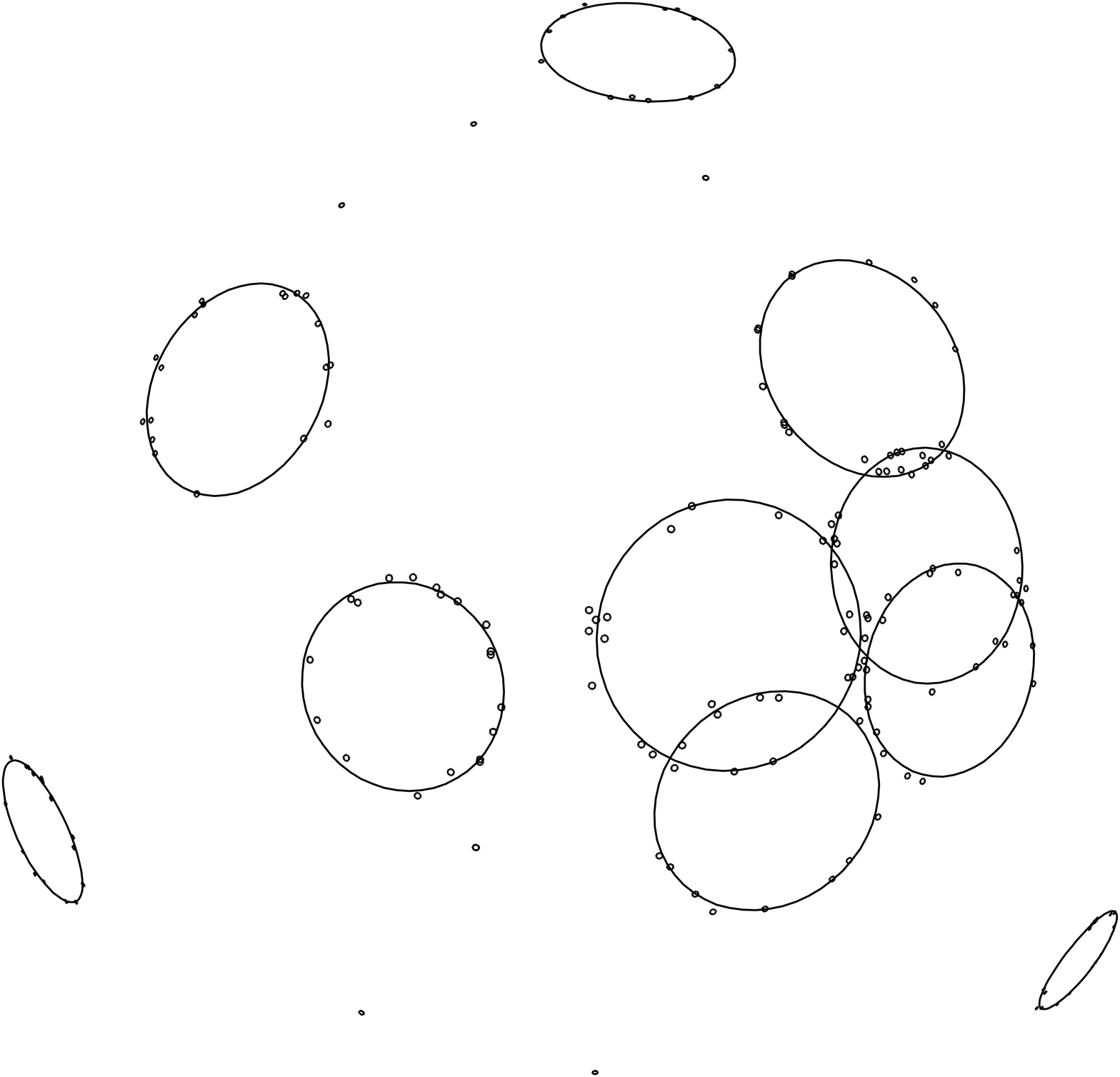,height=3cm}
}
\subfigure[event 4]{
\psfig{file=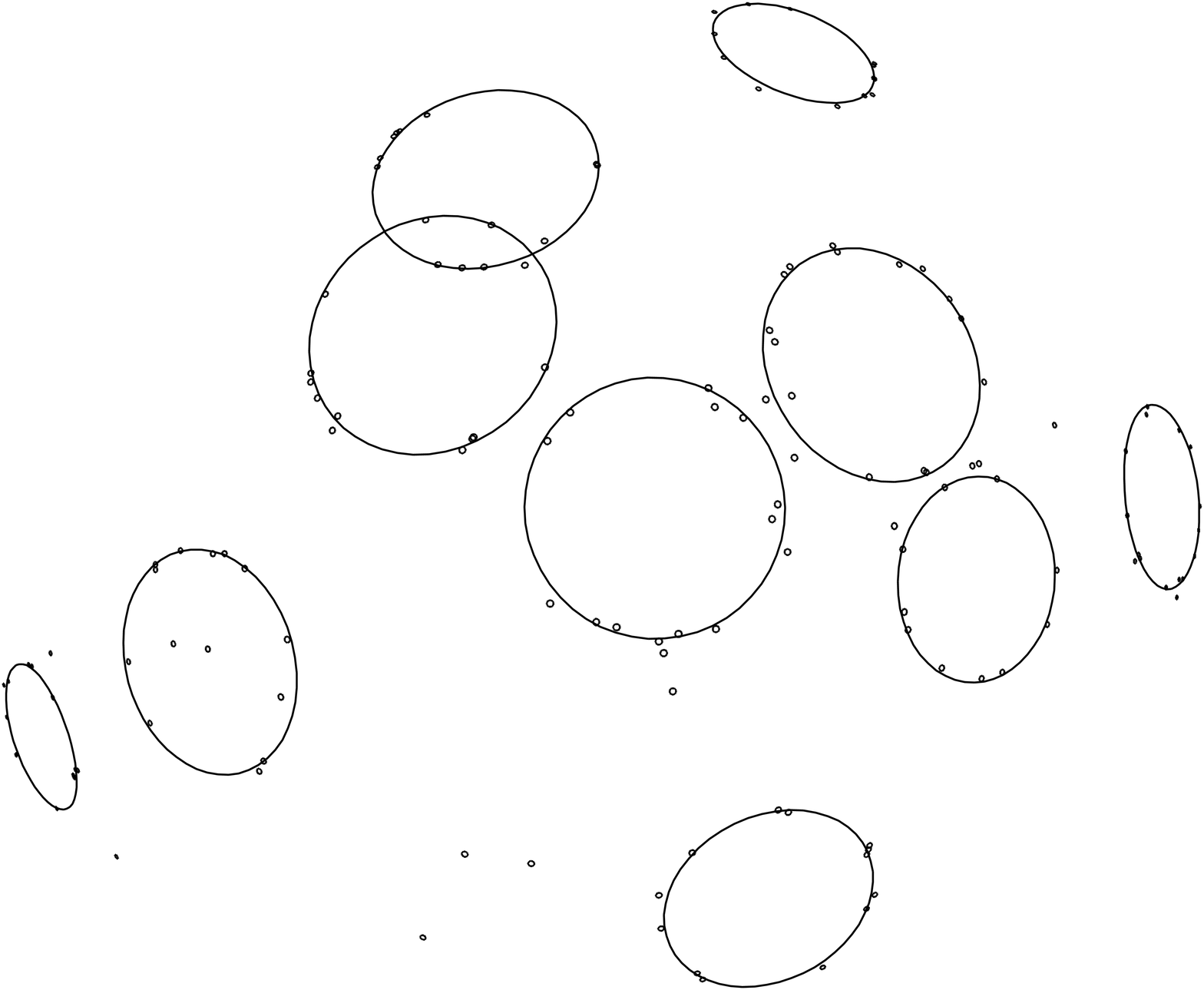,height=3cm}
\psfig{file=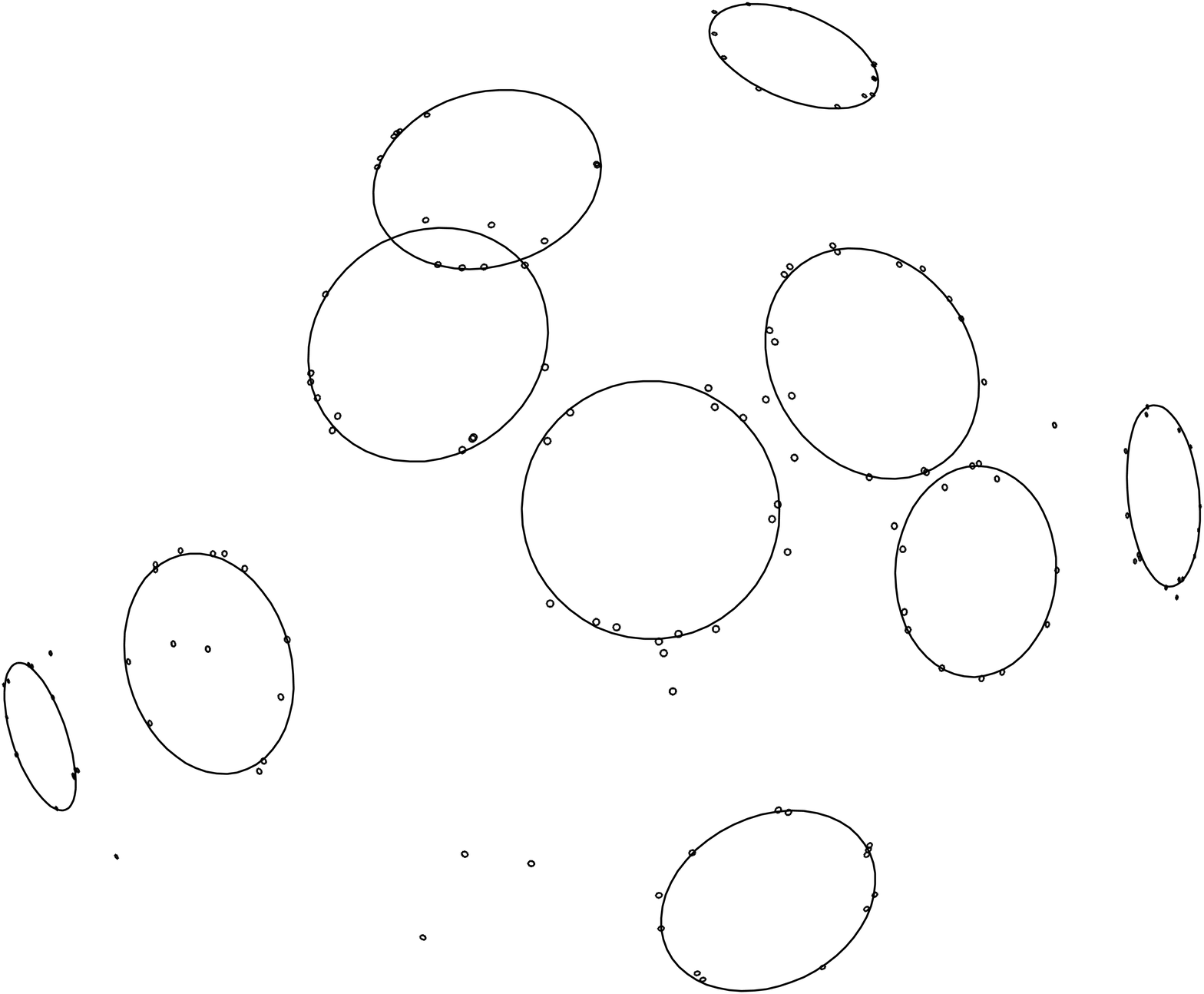,height=3cm}
}
\subfigure[event 5]{
\psfig{file=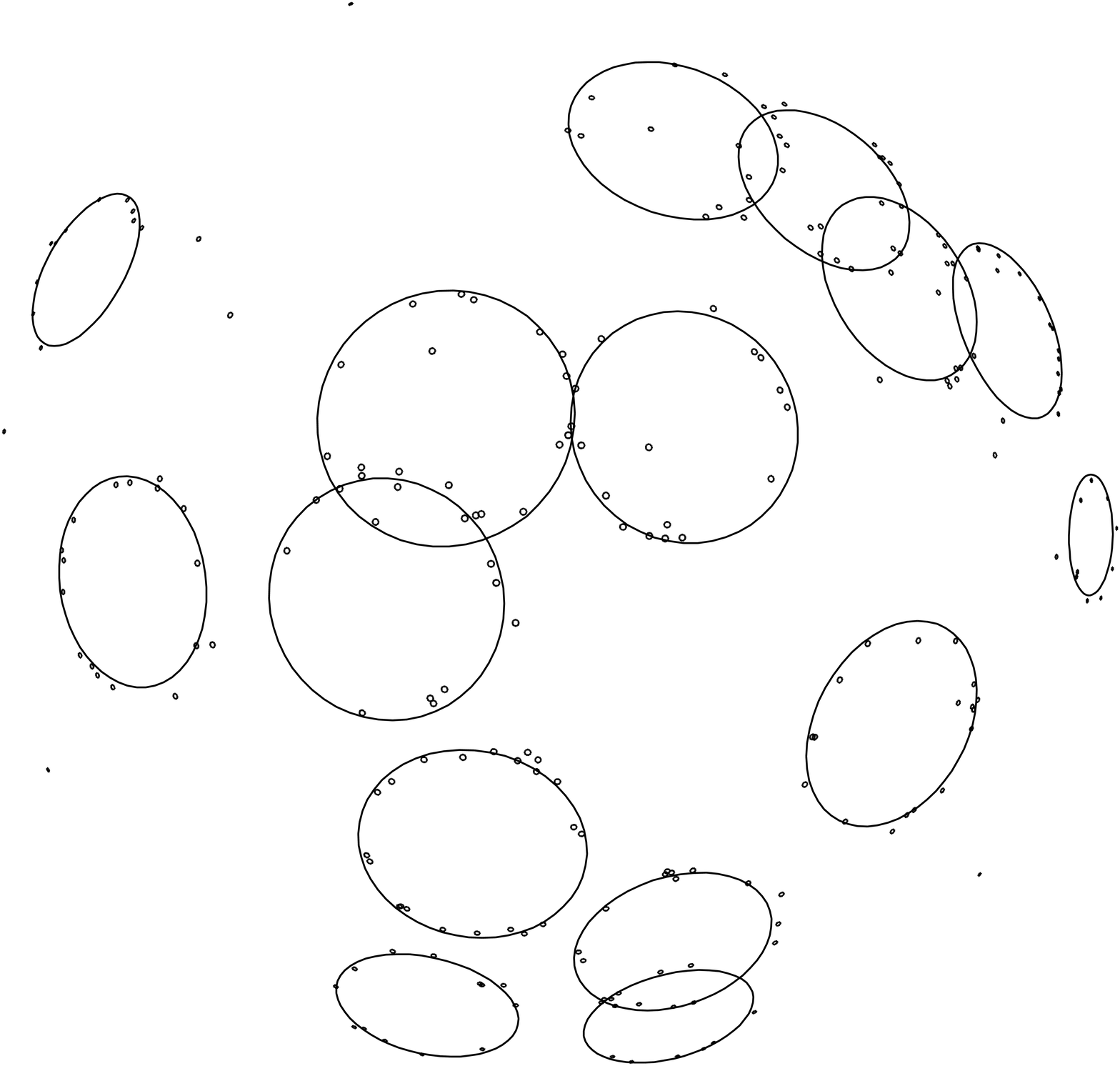,height=3cm}
\psfig{file=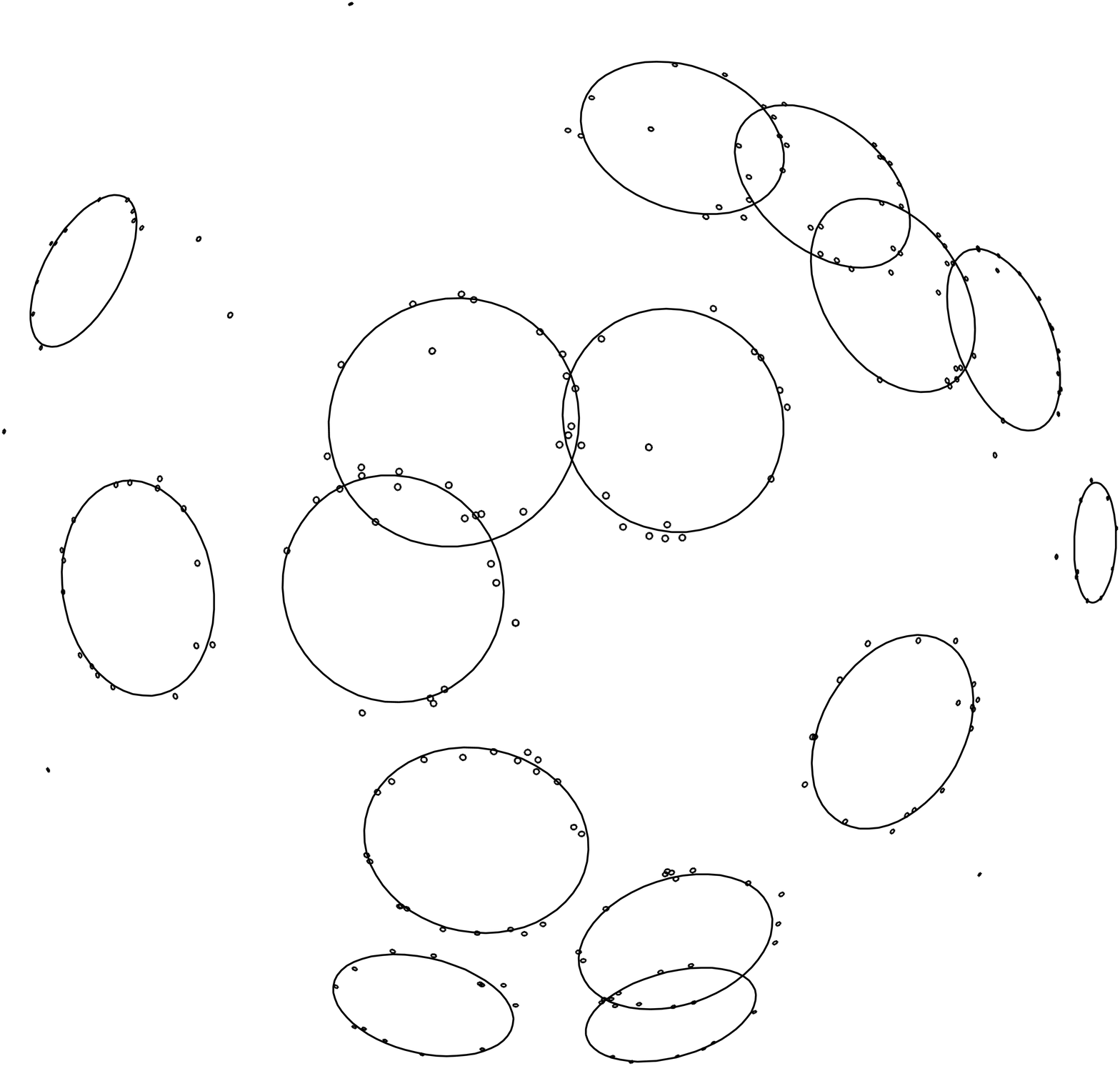,height=3cm}
}
\subfigure[event 6]{
\psfig{file=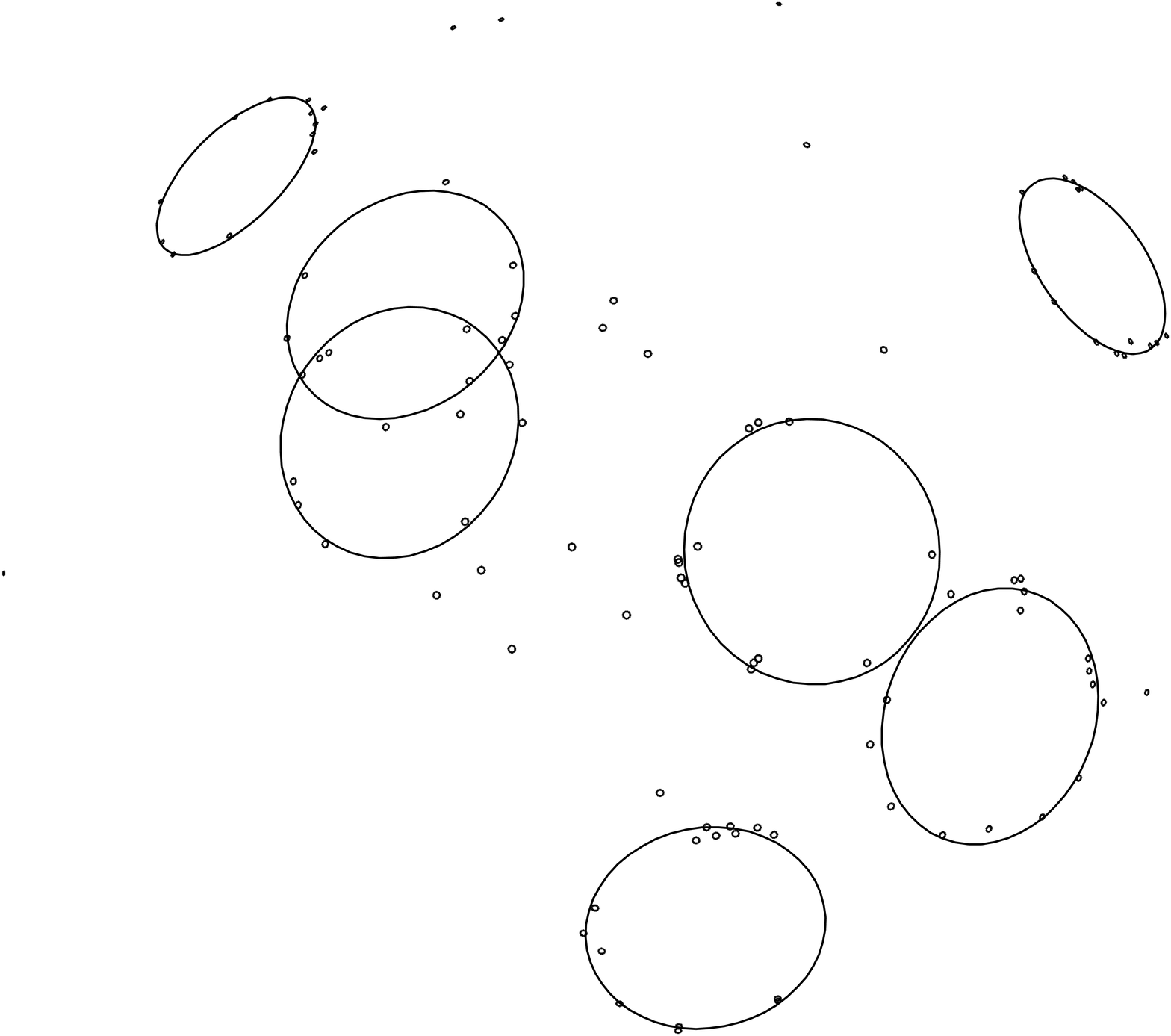,height=3cm}
\psfig{file=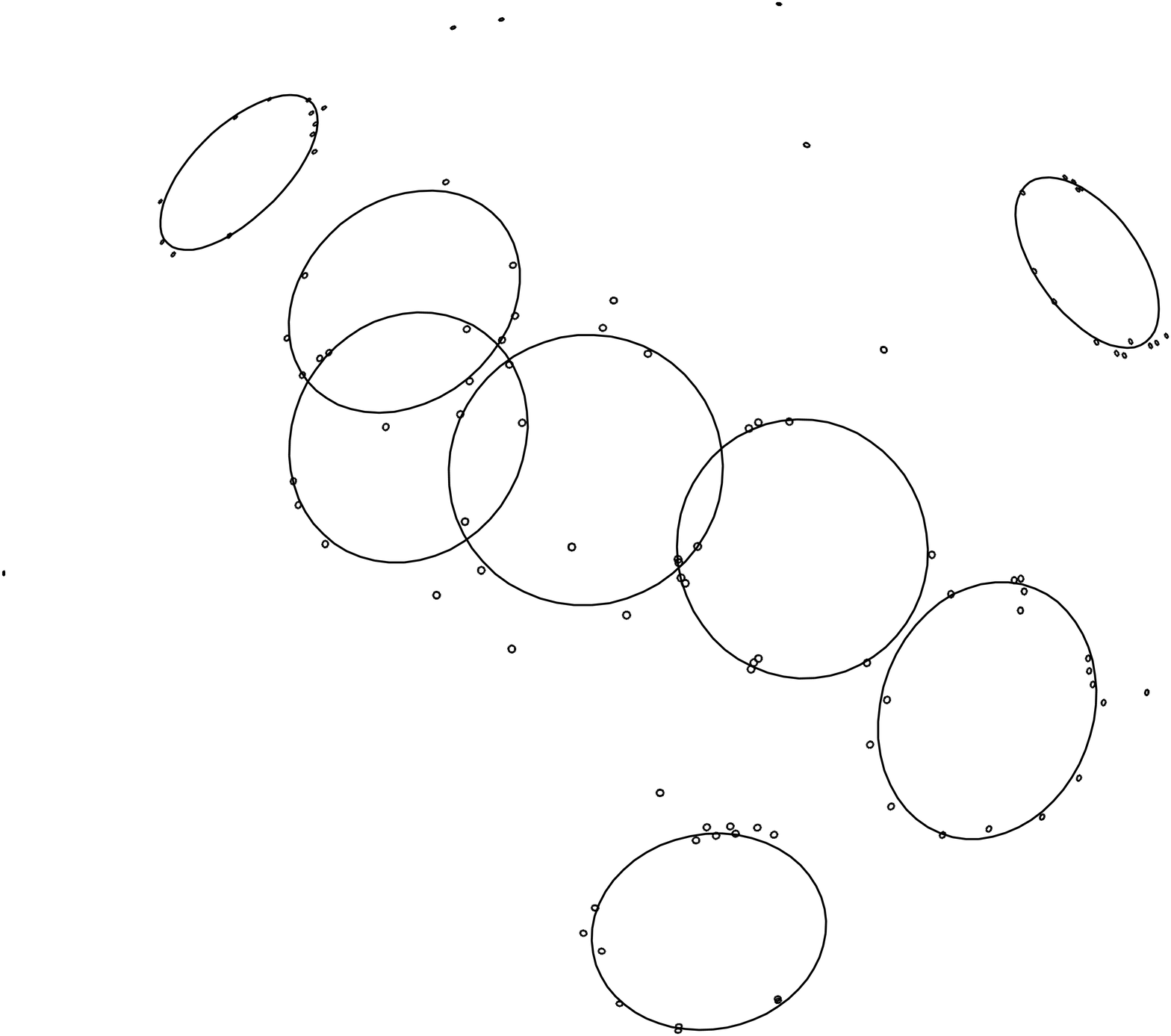,height=3cm}
}
\subfigure[event 7]{
\psfig{file=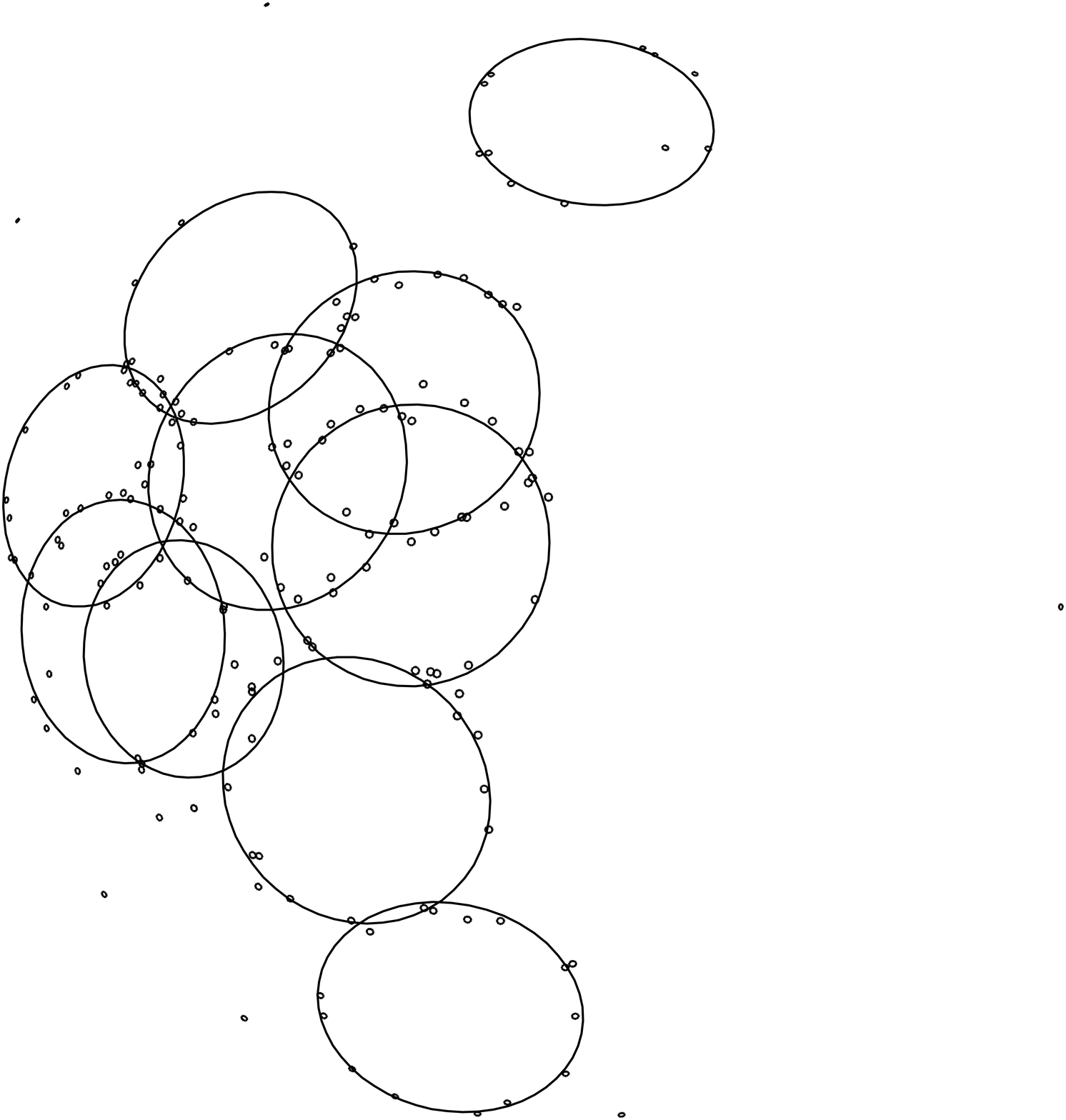,height=3cm}
\psfig{file=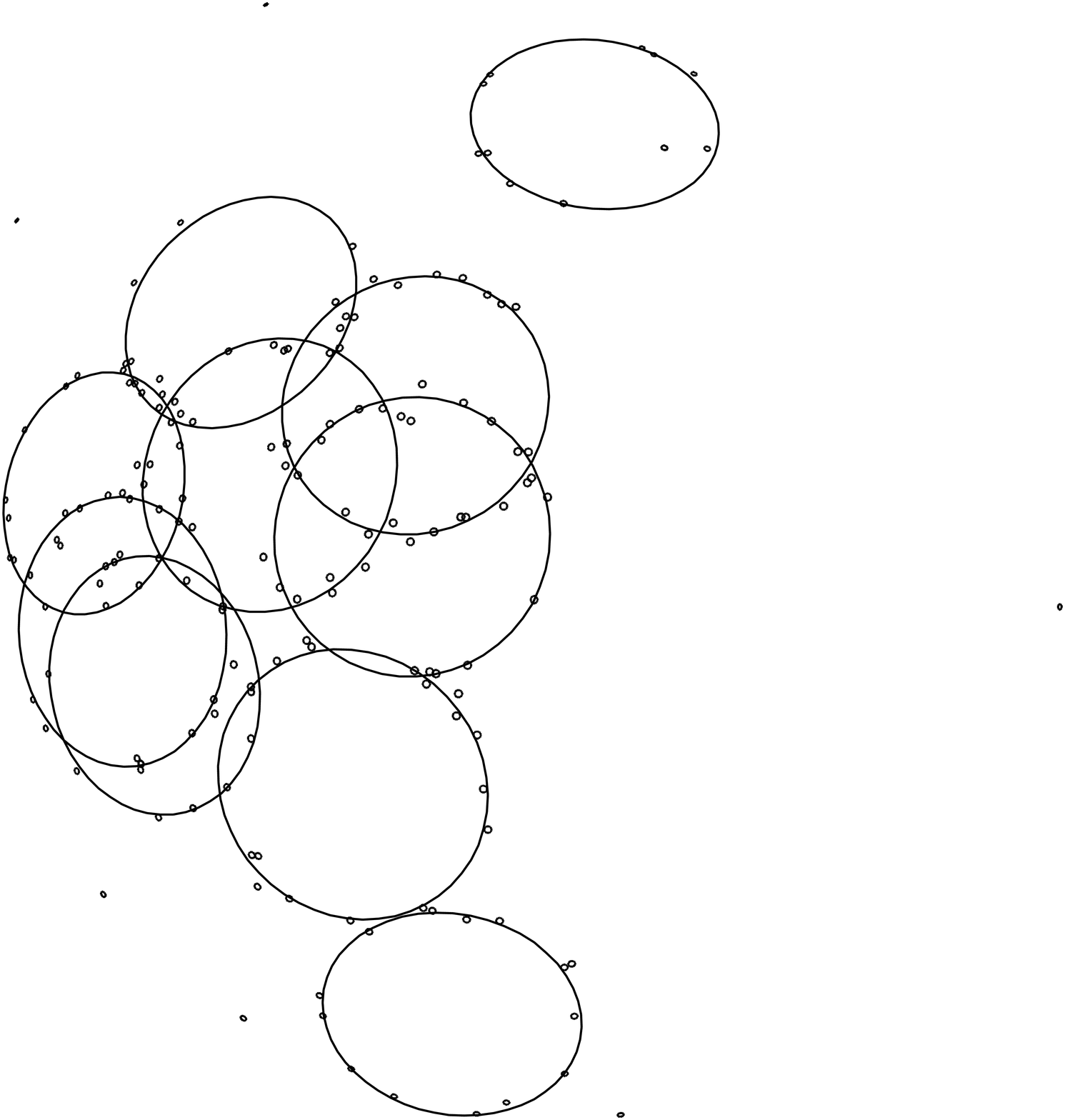,height=3cm}
}
\subfigure[event 8]{
\psfig{file=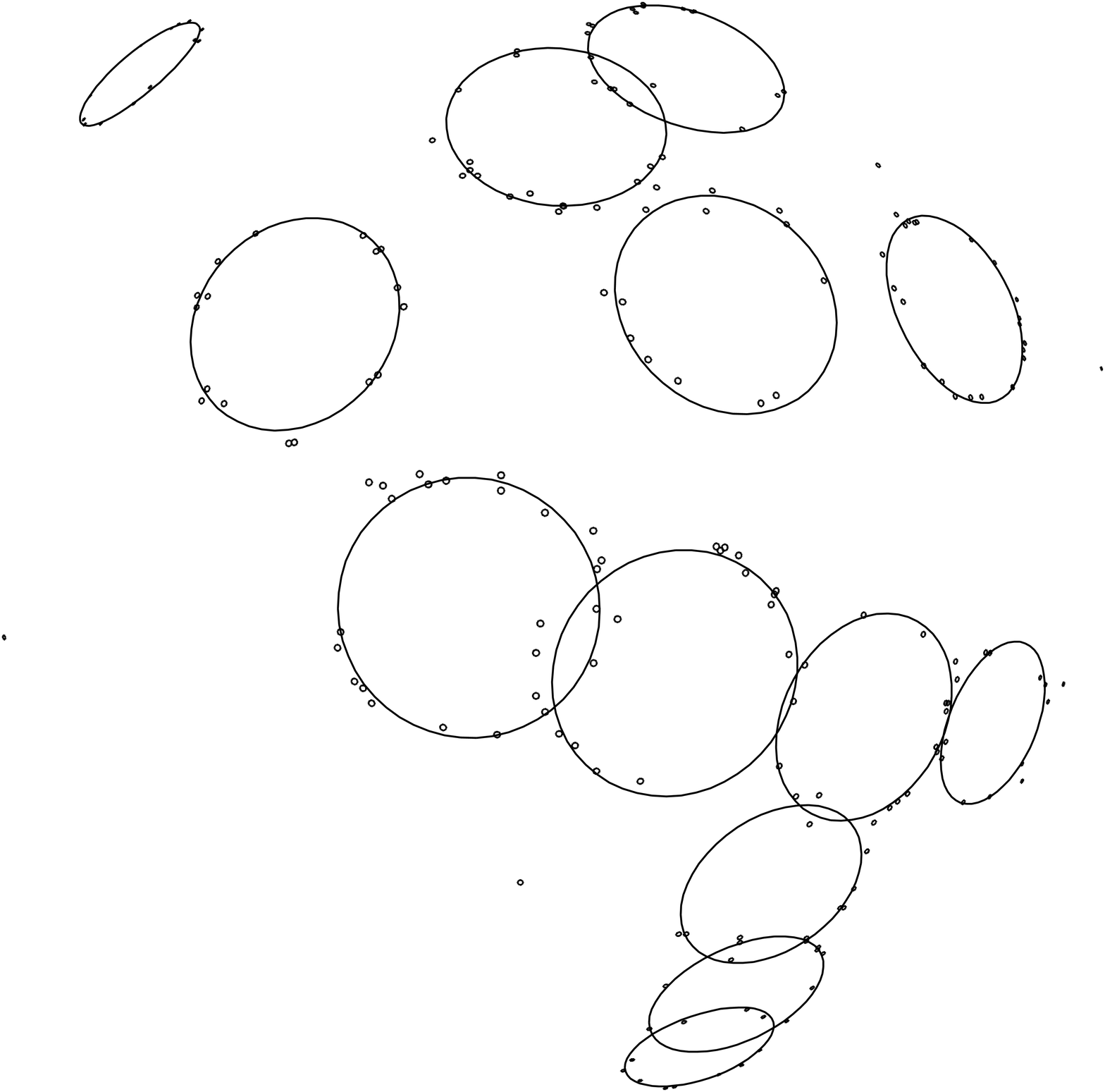,height=3cm}
\psfig{file=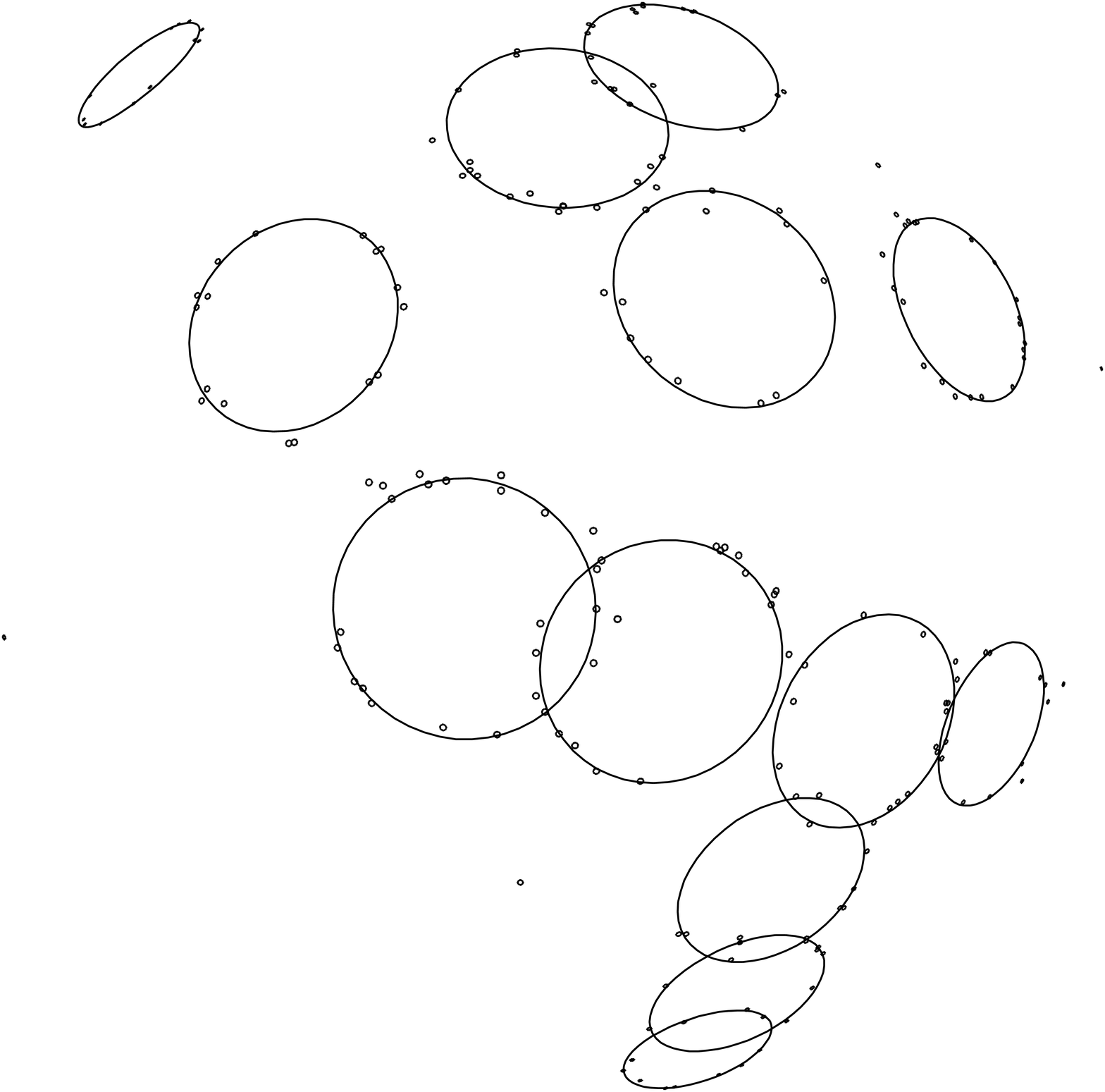,height=3cm}
}%
\caption{Fit performance allowing three seconds per event on a 3 GHz
  Pentium 4 computer.  In each event, the result of the fit (i.e.\ the sample
  taken 3 seconds after sampling began) is shown on the left, while
  the ``true'' distribution of rings which generated the hits is shown
  on the right.  All results are shown in a hyperbolic projection
  which compresses the whole of 2-space onto a disc.  This projection
  is the cause of the elliptic distortion at the periphery of the
  disc.
\label{fig:threeSecondFits}}
\end{figure}

\label{sec:secwithreconstructedringsinit}
 
Figure~\ref{fig:threeSecondFits} shows the fits obtained after three
seconds of sampling (each) for eight events on a 3 GHz Pentium 4
computer.  No special selection was applied to these events.  They are
just the first eight events generated according to the toy model of
Section~\ref{sec:notes}.  Note that in the time alotted the
\ringfinder\ missed a ring in event 6, and fitted a ring in event 7
poorly (the ring third from the bottom).  All other rings are fitted
well.  The lowest ring multiplicity observed in the eight events was 3
rings in event 1.  Event 5 had the most rings: 15.

More work needs to be done to optimise the termination criterion for
the sampling process. Just stopping each event after three seconds or
a fixed number of samples is very crude.  A more sensible stopping
criterion might choose to run complicated events (events with a large
number of hits) for longer than simple ones.  Even though better
stopping criteria may be found in the future, it is clear from the
very simple one implemented here that the \ringfinder\ is indeed able
to find rings.

\section{Conclusions}

This article has described an algorithm optimised for identifying
rings among photons in RICH detectors which are similar to the LHbC
RICH toy model of Section~\ref{sec:notes}.

The algorithm acts only on hits, and does not have to be seeded with
the locations of, for example, the centres of the rings.

The algorithm has demonstrated good performace on events produced by
the toy model described in the text, at a cost of 3-seconds per event
on a 3 GHz Pentium 4.

There is ample scope to optimise the \ringfinder\ further, by finding
better proposal functions and more realistic detector models.

\section{Acknowledgements}

The author would like to thank Dr. C.\ R.\ Jones for his continued
support throughout the development of the \ringfinder, and for his
comments on the draft document.  This work was partly funded by the
author's Particle Physics and Astronomy Research Council (PPARC)
Fellowship.

\section{Appendix: Details of Lester's Highly basic Computational
  (LHbC) RICH simulation}
\label{sec:notes}

This section lists the constituents of the toy model assumed for
hit-production in order to calculate explicit forms for $p_P({\bf
H}|{\bf R})$ and $p_M({\bf R})$ introduced in
Section~\ref{sec:introducingthedistros}.  
The same model was used to generate the events shown in
Figure~\ref{fig:threeSecondFits} which were subsequently fitted by the
\ringfinder.  It is hoped that the particular distributions and
numbers chosen to define Lester's Highly basic Computational (LHbC)
RICH simulation will make the events it generates similar to those
which might be seen in a future RICH detector of some kind.

The number of rings in an event was taken to be Poisson
distributed with mean 10.  The radius of each ring was assumed to be
distributed according to a probability distribution proportional to
the parameterization:
\begin{equation}
  {e^{625(x-0.0305)}} \over {
       (1 + e^{5(x-0.0305)})(1 + e^{2941(x-0.0305)}) }
\label{eq:radiusdistroeq}
\end{equation}
(for $x$ measured in radians) which is shown in
Figure~\ref{fig:ringRadii}.  The $x$ and $y$ co-ordinates of the centre of
each ring were taken to be independent and Gaussian distributed with
mean 0 and standard deviation 0.09 radians.  (All distance values are measured in radians as we work in angular
co-ordinates.)  The mean number of hits
per unit length $\rho$ on the circumference of each ring was
30/radian.  The actual number of hits on a ring with radius $r_R$ was
Poisson distributed with mean $2 \pi r_R \rho$.   The hits themseleves
were taken to be distributed uniformly in azimuthal angle $\phi$, and
with radii $r_h$ (distance from centre of ring) distributed
independently for each hit according to
\begin{eqnarray}
  p({\MarkovHitRadius | \MarkovRingRadius, \MarkovEpsilon,
    \MarkovAlpha})
 & \propto &
  \frac{1}{2 \pi \MarkovHitRadius^2}
  ({\frac{\MarkovHitRadius}{\MarkovRingRadius}})
  ^
  {\MarkovAlpha - 2}
  \exp(
    -\frac{\log^2\!(\MarkovHitRadius/\MarkovRingRadius)}
    {2\MarkovEpsilon^2} )
  \label{eq:YorkshirePuddingFunction}
\end{eqnarray}
in which the dimensionless parameter $\MarkovAlpha$ controlling the
distribution skew took the value 2, and in which the dimensionless
parameter $\MarkovEpsilon$ controlling the thikness of the ring
took the value 0.05.

The number of background hits (hits not coming from Cherenkov
photons) in an event was taken to be Poisson distributed with mean 10.
The $x$ and $y$ co-ordinates of each background hit were taken to be
independently distributed acoording to the same Gaussian distributions
used for the $x$ and $y$ co-ordinates of ring centres.


\bibliography{draft}

\end{document}